\documentclass[aip,twocolumn,superscriptaddress,prl,10pt]{revtex4-1}
\usepackage{amsmath,amssymb,graphicx,graphics}
\usepackage[colorlinks==true,pdfborder={0 0 0}]{hyperref}
\usepackage{setspace}
\usepackage{tabularx}
\usepackage{blindtext}
\begin{document}

\title{\large{Peierls transition and edge reconstruction in phosphorene nanoribbons}}

\author{Ajanta Maity}
\author{Akansha Singh}
\author{Prasenjit Sen}
\email{prasen@hri.res.in}
\affiliation{\small{Harish-Chandra Research Institute, Chhatnag Road, Jhunsi, Allahabad 211019, India.}}

\begin{abstract}
Atomic and electronic structures of phosphorene nanoribbons are studied within density
functional theory. These novel materials present different physical phenomena expected in two very different physical systems: 
one dimensional metallic chains and semiconductor surfaces.  While `rugged' nanoribbons are semiconducting in their layer-terminnated
structures, pure `linear' and `zigzag' nanoribbons are metallic due to metallic edge states. 
Linear nanoribbons undergo edge reconstruction and zigzag nanoribbons beyond a certain width undergo Peierls transition
leading to opening of a band gap in the electronic structure and lowering of total energy. Mixed nanoribbons with linear and zigzag edges
on the two sides turn out to be a curious case that display both edge reconstruction and Peierls transition simultaneously. Most phosphoeren
nanoribbons turn out to be semiconductors having important implications for their application.
\end{abstract}

\maketitle

Since the synthesis of graphene sheets~\cite{novoselov-04} there has been a steady interest in two dimensional (2D) materials
mainly for electronic applications. Graphene turned out to be a material with remarkable properties: very high charge carrier mobility
($\sim 200,000$ cm$^2$/V.s)~\cite{bolotin-08}, high thermal conductivity ($\sim$5000 W/mK) and unusually large mechanical strengths 
(Young's modulus of $\sim$2 TPa)~\cite{lee-12}. However, the advantage of high
mobility in field effect transistor (FET) devices is lost due to absence of a band gap in graphene. This leads to a large off-state current
and a low drain current modulation.

This has led to search for other 2D materials with reasonable mobilities and sizable gaps. A class of layered transition metal dichalcogenides (TMD's),  
MoS$_2$, MoSe$_2$, WS$_2$ and WSe$_2$, have attracted a lot of attention lately~\cite{chhowalla-13}. Mono- and few-layers of these materials have
been synthesized through mechanical and chemical exfoliation~\cite{wang-12}. These have shown mobility values of a few hundred cm$^2$/V.s. 
Drain current modulation as high as 10$^8$ has been achieved in FET devices made of a monolayer MoS$_2$~\cite{radisavljevic-11}.

Very recently, black phosphorus (BP), the most stable allotrope of phosphorus, has attracted great attention. BP is a layered material like
graphene in which successive layers are held together by van der Waals (vdW) forces~\cite{vaithee}. Few-layer samples of phosphorene (name given to a
monolayer of BP) have been exfoliated using mechanical means~\cite{li-14,liu-14,gomez-14,buscema-14}. FET devices have also been made using few-layer phosphorene as
the channel. Few-layer phosphorene offers a reasonable mobility and band gap. While the measured band gap of bulk BP is 
$\sim 0.3$ eV~\cite{warschauer-63,narita-83,maruyama-81}, it increases
with decreasing number of layers. The calculated band gap of a monolayer 
is $\sim 1.6$ eV~\cite{liu-14}. Mobility as high as $\sim 1000$ cm$^2$/V.s 
has been measured in few-layer devices and a drain current modulation of 10$^5$ has been achieved~\cite{li-14}. These results make phosphorene an attractive
material for electronic applications and one needs to understand its fundamental physical properties well.

A few theoretical studies have already been reported on mono-layer and few-layer phosphorene, and phsphorene nanoribbons (PNR).
Apart from the band structure of mono- and few-layer phosphorene~\cite{li-14,liu-14,tran-14}, electron and hole effective mass and charge carrier
mobility have been calculated within the density functional theory (DFT)~\cite{qiao-14}. Mechanical~\cite{wei-14} and elastic properties~\cite{jiang-14} and strain engineering of 
band gaps~\cite{rodin-14,peng-14} have also been explored. Guo et al~\cite{guo-14} have studied PNR's, phosphorene nanotubes and
vdW multilayers with TMD's within density functional theory (DFT). They reported band structure of two different types of PNR's one of which was found to be metallic
and the other semiconducting.

In this letter we study atomic and electronic structure of PNR's in detail and find these to be a playground for interesting physics
usually expected in two very different classes of systems: one dimensional (1D) metallic chains and semiconductor surfaces. 
PNR's can have three different types of edges.  We study four types of nanoribbons, three are `pure' having both edges of the same type, and the
fourth being a mixed PNR having two different types of edges.
One type of pure PNR's are found to be indirect gap semiconductors for all widths studied so far. 
A second type of pure PNR's undergo edge reconstruction much as found on semiconductor surfaces,
while the third type of pure PNR's beyond a certain width undergo Peierls transition expected in 1D metallic chains. 
We have studied mixed PNR of only one width and it displays edge reconstruction at one edge and Peierls transition at the other.

\begin{figure}
\scalebox{0.35}{\includegraphics{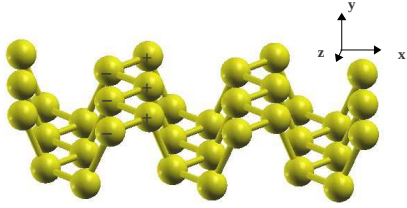}}
\caption{Structure of a monolayer phosphorene. See text for explanation of atoms marked by `$+$' and `$-$' signs.}
\label{fig:structure-ML}
\end{figure}

All our DFT calculations are done with 
an energy cutoff of 500 eV for the planewave basis set. Interactions between the valence electrons and the ion cores are represented by projector 
augmented wave (PAW) potentials. The PBE gradient corrected functional~\cite{pbe} is used for the exchange-correlation energy. In a couple of cases
we have also used the hybrid HSE06 functional~\cite{hse1,hse2}.  
An ($8\times 1\times 8$)  Mokhorst-Pack (MP) k-point mesh is used for calculations of the monolayer. For the nanoribbons
running along the $x$ and $z$ directions (defined below) ($8\times 1\times 1$) and ($1\times 1\times 8$) MP k-point meshes were employed.
We kept a vacuum space of 15 \AA~ in the non-periodic directions in the supercell. All the atoms in the supercells were relaxed using a conjugate gradient 
method till all the force components became less than 0.01 eV/\AA. The VASP code~\cite{vasp1,vasp2,vasp3,vasp4} was used for all the calculations. 
Atomic structure of a monolayer of phosphorene is shown
in Fig.~\ref{fig:structure-ML}. The direction along the trenches is designated as $z$ in this work, while the perpendicular direction in the plane of the monolayer
is designated as $x$. The direction perpendicular to the monolayer plane is designated as $y$. The optimum lattice constants of a monolayer 
along the $x$ and $z$ directions are found to be 4.62 \AA~ and 3.3 \AA~ respectively in good agreement with other DFT calculations~\cite{qiao-14}.

\begin{figure}
\scalebox{0.35}{\includegraphics{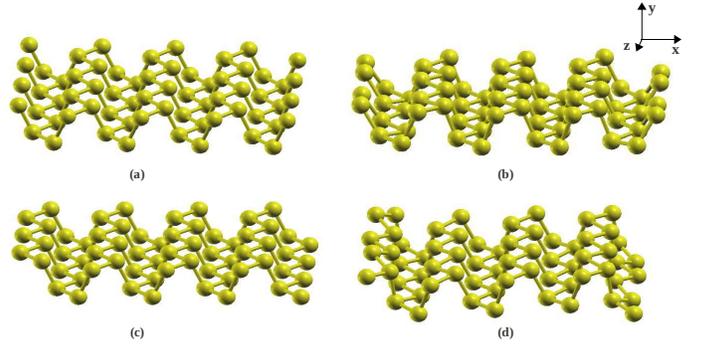}}
\caption{Structures for (a) layer-terminated l-PNR, (b) edge reconstructed l-PNR, (c) layer-terminated z-PNR, 
(d) Peierls distorted z-PNR. Both these are periodic along $z$ and finite along $x$.}
\label{fig:nr-lt}
\end{figure}

A phosphorene monolayer turned out to be a direct gap semiconductor with a gap of 0.9 eV in PBE and 1.6 eV in HSE06, in very good agreement 
with the values reported earlier~\cite{li-14,qiao-14}. Although a monolayer
phosphorene is a semiconductor with a sizable gap, our ability to tune the gap would give us the flexibility of using it in different applications.
Confinement is a standard way of tuning band gaps that has been
explored extensively in the context of graphene nanoribbons ~\cite{ezawa-06,son-06a,barone-06}. 
Therefore confining phosphoerene further along one direction may allow us to tune the gap, or it could even be metallic. 
This is our motivation for studying PNR's. One can form PNR's that are periodic along the $z$ direction and constrained along $x$  
or vice-versa. In a monolayer phosphorene, each
P atom is $sp^3$ hybridized. Each atom is covalently bonded to three neighbors, and has one lone pair of electrons in the fourth orbital~\cite{rodin-14}. 
A careful inspection
of Fig.~\ref{fig:structure-ML} shows that one may have two fundamentally different types of edges for PNR's that are periodic along $z$.  
For example, the left edge may be formed either by the atoms marked $+$ or by those marked $-$ in Fig.~\ref{fig:structure-ML}. In case of the former, in the layer-terminated
edge, each P atom is bonded to only one atom in the interior of the nanoribbon, 
and has two dangling bonds. In the latter type, each edge atom has only one dangling bond and two neighbors in the interior. We will call these the
linear and zigzag edges, and pure PNR's with these edges the linear PNR and zigzag PNR respectively (l-PNR and z-PNR).  
Structures of pure l-PNR and z-PNR are shown in Figs.~\ref{fig:nr-lt}(a) and (c). 
It can also be seen from Fig.~\ref{fig:structure-ML}  that in
case of PNR's periodic along $x$, only one type of edge is possible in which of the four atoms forming the unit cell in the periodic direction
two are 2-fold coordinated and the other two are 3-fold coordinated. We call these
the rugged edges and these PNR's the rugged nanoribbons (r-PNR). Structure of a r-PNR with layer terminated edges is shown in Supplementary Figure S1(a)~\cite{suppl}.
We mainly focus on pure of all three types of widths between 1-6 unit cells (of the monolayer), and a few other sizes as discussed below. For the mixed PNR, we took a pure l-PNR
of width 8 and removed the first row of atoms on one of the edges to create a zigzag edge.

First we discuss our results for r-PNR's. The narrowest ribbon thus has a width of one unit cell which in this case is a rugged chain of P atoms. 
This turns out to be a semi-metal in PBE with the valence and conduction bands touching at the $\Gamma$-point. But this is an artifact of the PBE 
functional as gradient corrected functionals are known to underestimate band gaps~\cite{kahanoff}. HSE06 gives a semiconductor with a direct gap of 0.45 eV at the $\Gamma$-point.
There is a significant distortion to the structure. All the atoms
nearly become co-planar in a plane containing the periodic direction with all the P-P bonds being $\sim 2.2$ \AA.
The relaxed structure of this PNR is shown in supplementary Figure S1(b). 
r-PNR's of width of 2-6 unit cells turn out to be indirect band gap semiconductors even in PBE.
For widths 2 and 3 the conduction band minimum (CBM) occurs at the $\Gamma$-point while the valence band maximum (VBM) is at a 
k-point intermediate between $\Gamma$ and
the zone boundary. For r-PNR's of widths 4-6, the VBM appears at the $\Gamma$-point and the CBM appears between the $\Gamma$ and the zone boundary.
The band gap increases as the width increases from 2 to 3, and then decreases up to width 6. We find a gap of 0.45 eV for a r-PNR of width 6.
Guo et al~\cite{guo-14} have studied these nanoribbons (called armchair PNR's) of width 7-12 unit cells and
find them to be indirect gap semiconductors. We calculated a ribbon of width 10 to have an overlap with the range of widths 
studied by these authors. This also turned out to be a semiconductor having an indirect gap of 0.41 eV in agreement with the
results in ref.~\cite{guo-14}. Calculated band gaps of all the semiconducting PNR's are given in the Supplementary Information~\cite{suppl}.
One may conclude that the band gap of r-PNR's is not very sensitive to the width beyond 6 unit cells.
Band structure plots for nanoribbons of widths 1, 2 and 3 unit cells are shown in  supplementary Figure S2. 
Thus an indirect to direct gap transition occurs somewhere between a ribbon of width 12 unit cells and the monolayer.

\begin{figure}
\scalebox{0.35}{\includegraphics{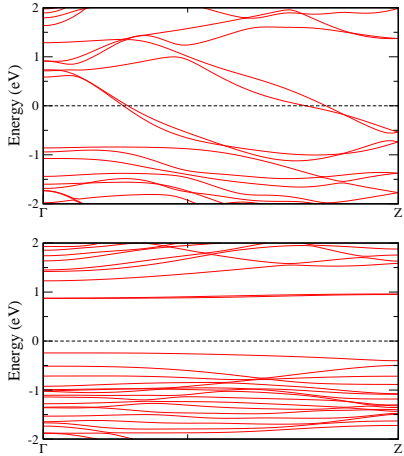}}
\caption{Band structures of a 4 unit cells wide (a) layer-terminated l-PNR, (b) edge reconstructed l-PNR.}
\label{fig:bs-lpnr}
\end{figure}

\begin{figure}
\scalebox{0.5}{\includegraphics{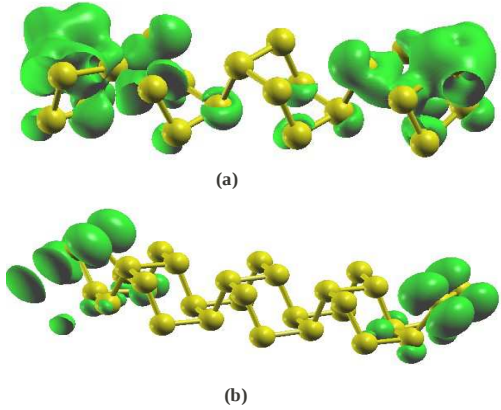}}
\caption{(a) Charge density corresponding to one of the bands crossing the Fermi energy in a layer-terminated l-PNR. (b) Charge density from
the lowest conduction band in edge reconstructed l-PNR.}
\label{fig:parchg}
\end{figure}

Now we focus on pure l-PNR's and z-PNR's that turn out to be more interesting from a fundamental perspective. We calculated
band structure of l-PNR's of widths ranging from 1-6, and 8 unit cells taking the same unit cell in the periodic direction as in a 
monolayer. We call these the layer-terminated PNR's in contrast to structurally distorted PNR's that have different unit cells in the
periodic direction (discussed below). 
A l-PNR of width 1 turns out to be a semiconductor with an indirect gap
of 1.38 eV. The CBM is at the $\Gamma$-point and the VBM is between $\Gamma$ and the zone boundary. 
l-PNR of width 2 in its layer-terminated structure presents and interesting case where two bands cross each other at the Fermi energy.
These band structure plots are shown in supplementary Figure S3. Ribbons of widths 3-6 turn out to be metallic with four 
bands crossing the Fermi energy in each case. Two of these bands are $\sim 1/4$ filled and two others are $\sim 3/4$ filled. 
Band structure for a l-PNR of width 4 is shown in Fig.~\ref{fig:bs-lpnr}(a). The topmost valence band and the lowest conduction band
in the monolayer have mainly $p_y$, and $p_y + p_x$ characters respectively.
The bands crossing the Fermi energy have major contributions from the $p_x$ and $p_z$ orbitals of the edge atoms only. So clearly, these
are edge states with different bonding characteristics than states in the monolayer. These originate from $\sigma$ overlap of the $p_z$ orbitals, and $\pi$ 
overlap of the $p_x$ orbitals on the neighboring atoms and form metallic channels running along the edges. 
A charge density isosurface plot for one of these bands (Fig.~\ref{fig:parchg}(a)) clearly 
show that it is localized at the edges in the transverse direction but is delocalized along both the edge channels. Other
bands crossing the Fermi energy give similar pictures.

This metallic band structure is similar to those obtained in many bulk terminated semiconductor surfaces. Such metallic 
surface states result within the bulk band gap due to weak overlap between orbitals of nearest neighbor surface atoms that are actually
next nearest neighbors in the bulk. This is exactly what happens in the l-PNR's. Successive edge atoms in the periodic direction are in fact next nearest neighbors in
the monolayer having a separation of 3.3 \AA~ compared to the nearest neighbor distance of $\sim 2.2$ \AA. It is well known that many semiconductor
surfaces, for example, C(100), Si(100), Ge(100), undergo
reconstructions through dehybridization of the $sp^3$ hybrid orbitals on the surface atoms and formation of new bonds~\cite{bechstedt}. 
The end result is the formation of surface
dimers that leads to doubling or quadrupling of the surface unit cell. The natural question to ask is whether such an edge reconstruction is
possible in the l-PNR's. To test this, we studied l-PNR's of widths 1-6 and 8 taking supercells having two primitive cells along the periodic $z$ direction.
After atomic relaxation, the two neighboring atoms on each edge move towards each other and form an edge dimer. The dimer bond length is only 
$\sim 2$ \AA, even smaller than the nearest neighbor bond length in the monolayer. This is a consequence
of the fact that each edge P atom makes only one bond in the layer-terminated structure which gives it a large degree of structural flexibility.
Hence, there is a significant reconstruction at the edges with a consequent doubling of the unit cell along the periodic direction. 
This indeed opens a gap in the band structure and leads to a reduction in total energy. The gap in a 4 unit cell wide l-PNR is
1.11 eV (direct gap) and the reduction in total energy is 1.44 eV per edge dimer. 
The energy gain due to edge reconstruction is substantial when compared to the energy gain in the well studied $(2\times 1)$ reconstruction
of the Si(100) surface ($\sim 2$ eV per surface dimer)~\cite{ramstad-95}. The band structure of an
edge reconstructed l-PNR of width 4 is shown in Fig.~\ref{fig:bs-lpnr}(b). This is the result of a dehybridization 
process on the edge atoms. The edge bands crossing the Fermi energy are formed of the $p_x$ and $p_z$ orbitals as already stated. 
But after reconstruction, two degenerate CBM's are formed by the $p_x$ and $p_y$ orbitals of the edge atoms. These
bands have very little dispersion because these are localized on the dimers and have very little amplitude in between two successive edge dimers. 
This can be understood from the fact that the smallest separation between two successive edge dimers is 4.6 \AA. 
We have illustrated the difference between the metallic edge bands and the CBM after edge reconstruction in Fig.~\ref{fig:parchg}.
The l-PNR of width 2, which showed a crossing of bands, also undergoes the same edge reconstruction that opens an indirect 
gap of 1.31 eV as found in PBE. 

\begin{figure}
\scalebox{0.35}{\includegraphics{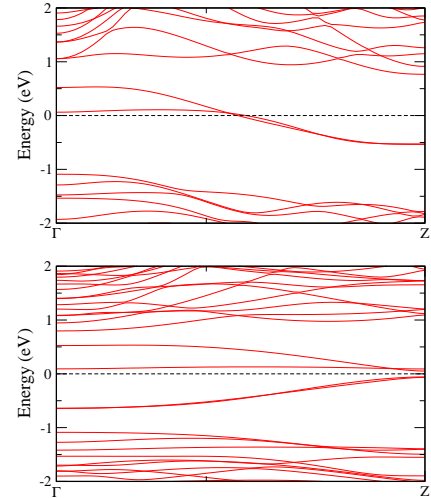}}
\caption{Band structures of a 4 unit cell wide (a) layer-terminated z-PNR, (b) Peierls distorted z-PNR. }
\label{fig:bs-zpnr}
\end{figure}

z-PNR's are also found to have metallic band structure at all widths between 1-6 in their layer-terminated structures. 
However, the electronic structure is distinctly different from that of the l-PNR's. In z-PNR's of all widths only two bands cross the Fermi energy.
In nanoribbons of width larger than 3 unit cells, both these bands are exactly half-filled. This is understandable because there is only one dangling 
bond per edge atom in the z-PNR's. The band structure turns out
to be different in z-PNR's of width 1-3. The k-points at which these bands cross the Fermi energy depends on the width of the PNR. However,
none of these crossings happens at simple fractions such a 1/2 or 1/4 from $\Gamma$ to the zone boundary. For example, one of the
bands in the z-PNR of width 1 crosses the Fermi energy at 9/10 to the zone boundary.
Focussing on the PNR's of width 4 and more, our calculated band structures are similar to those of Guo et al~\cite{guo-14}. 
We also studied a z-PNR of width 8 to have an overlap with ref.~\cite{guo-14}, and find exactly the same band structure.
Band structure of a layer-terminated z-PNR of width 4 is shown in Fig.~\ref{fig:bs-zpnr}(a).
The bands crossing the Fermi energy are formed of the $p_x$ and $p_y$ orbitals of the edge atoms.
Just as in case of the l-PNR's, one can ask whether a reconstruction is possible in this case also.
To test this we again calculated the electronic structure of z-PNR's taking two unit cells in the supercell. z-PNR's of width 1-3 retain their 
atomic structure and metallic character. However, interesting changes in the structure and
consequently the electronic structure are found for the wider z-PNR's. The distance between successive atoms in the periodic direction
along the first, second and third rows of atoms starting from the edges
decreases marginally to 3.27 \AA~ (from 3.3 \AA) in the relaxed structure. Beyond these, in the interior of the PNR, the atoms have a separation of 3.3 \AA, as in a monolayer.
This small distortion is enough to open a gap of $0.11$ eV at the zone boundary (for width 4), and decreases the total energy by a small amount of 0.6 meV/atom. 
Although this energy change is at the limits of DFT calculations, the gap in the band structure is significant because PBE underestimates the gap value.
Band structure of a z-PNR of width 4 after gap opening is shown in Fig.~\ref{fig:bs-zpnr}(b).
A small structural distortion along the periodic direction leading to a doubling of the unit cell, a small decrease in energy and opening of a small gap 
at the zone boundary are reminiscent of Peierls transition in one-dimensional metallic chains~\cite{peierls}.  If this
is indeed the case, one expects the atomic character of the VBM and the CBM after gap opening to be
the same as the character of the edge bands crossing the Fermi energy in the layer-terminated metallic nanoribbons. Indeed, the VBM at the zone boundary
is found to be formed of the $p_x$ and $p_y$ orbitals of the edge atoms in the Peierls distorted PNR. The CBM, however, have contributions from some of the
interior atoms in addition to the edge atoms. The reason z-PNR's of width 1-3 do not undergo Peierls transition in our calculations with supercells containing
two unit cells in the periodic direction is not difficult to understand.
Since the bands responsible for their metallic character do not cross the Fermi energy at the half-way point from $\Gamma$ to the zone boundary,
a doubling of the unit cell does not open a gap. It is possible that more complicated structural reconstructions involving more atoms in the
periodic direction may fold the Brillouin zone appropriately so that a gap opens up at the zone boundary. However, it is difficult to guess
such distortions, particularly with two bands crossing the Fermi energy at two different k-points.

Lastly we discuss our results for the mixed PNR. In the layer terminated structure it is found to be metallic with three bands crossing the Fermi energy.
Two of these bands have major contributions from the linear edge and the third have major contributions from the zigzag edge. However, unlike pure
PNR's of similar width, none of these bands are pure edge states and they have non-negligible contributions from the interior atoms as well. This is a 
consequence of the lack of inversion symmetry in the transverse direction in a mixed PNR. In pure PNR's of both types, each band crossing the Fermi
energy has equal weights on both the edges. That is not allowed in the broken symmetry mixed PNR's. Therefore, a state having large amplitude at the
zigzag edge has to die out as one moves across the width of the PNR towards the linear edge. However, the amplitude does not vanish abruptly as
one moves away from the edge, and has finite values on some interior atoms as well. The same argument applies to states having large amplitude on
the linear edge. There are two bands having large amplitudes on the linear edge and one band on the zigzag edge because
each atom on the former has two dangling bonds and on the latter has one dangling bond. Band structure plot for the mixed PNR is shown in supplementary 
Figure S4. The bands do not cross the Fermi energy at simple fractions of the distance between $\Gamma$ and $Z$ though together they accommodate
three electrons. Therefore, it is not clear that a simple dimerization at the edges would lead to opening of a gap. Yet we tried this, and found that the
linear edge undergoes a reconstruction and the zigzag edge shows Peierls transition exactly as they do in the pure PNR's. The dimer bond length at the
liner edge is $\sim 2$ \AA, exactly as in the pure l-PNR and the first three rows of atoms at the zigzag edge reduce their bond lengths by the
same amount (0.03 \AA) as in a pure z-PNR. And interestingly and unexpectedly, this opens up a small gap in the band structure (Figure S4) that is 0.125 eV. The total
energy is lowered by 1.44 eV. This is same as lowering of energy per dimer in pure l-PNR's because the major contribution to the energy gain
comes from the linear edge, lowering due to Peierls transition at the zigzag edge being very small. After reconstruction and Peierls transition, the
two lowest conduction bands are formed by the $p_x$ and $p_y$ orbitals on the linear edge atoms. The top-most valence band has major contributions
from the $p_x$ and $p_y$ orbitals of the zigzag edge atoms. These are similar to what were found on the pure PNR's. A mixed PNR, therefore, turns out to be
a rare example that undergoes both edge reconstruction and Peierls transition simultaneously.

We now present a picture of relative stability of various PNR's. In their layer-terminated structures, z-PNR's turn out to be the most stable with a cohesive
energy of 5.323 eV per atom.
Cohesive energy per atom is define as $E_c = (nE_{\rm P}- E({\rm nPNR}))/n$. Where $ E_{\rm P}$ is the energy of an isolated
P atom and $ E({\rm nPNR})$ is the total energy of a PNR containing $n$ atoms in the supercell. z-PNR is followed by the l-PNR having a cohesive 
energy of 5.290 eV per atom. The r-PNR is the least stable with a marginally lower cohesive energy of 5.288 eV per atom. After edge reconstruction, the l-PNR has
a cohesive energy of 5.333 while the z-PNR has a cohesive energy of 5.323 eV per atom after Peierls transition. The mixed PNR has a cohesive
energy of 5.305 eV per atom in the layer terminated structure. This value is in between those of the l-PNR and the z-PNR, as one would expect.
After edge reconstruction and Peierls transition it has a cohesive energy of 5.327 eV per atom, again between those of l-PNR and z-PNR.
Thus edge reconstructed l-PNR happens to be the lowest energy structure. These numbers are for PNR's of width 8. 

In summary, our first-principles electronic structure calculations based on DFT show that
all PNR's except for z-PNR's of 1-3, are semiconducting. It is possible that these PNR's also become
semiconducting with complicated structural deformations. However, it is difficult to guess the exact nature of such structural deformations.
Experimental investigations can reveal the nature of atomic and electronic structure of these nanoribbons.
Ref.~\cite{guo-14} reported a metallic character for z-PNR's because they did not consider the possibility of a Peierls distortion along the edges. As we have
clearly demonstrated, l-PNR's undergo edge reconstruction and z-PNR's undergo Peierls distortion to open gaps in their band structures. Mixed PNR's
turn out to be a remarkable class of materials that show both edge reconstruction and Peierls transition.  These phenomena make
PNR's an interesting class of materials from a fundamental point of view. This information will also be useful for any electronic applications of this material.
We hope that these results will motivate experimental search for these effects in PNR's.

{\bf Acknowledgement} We acknowledge useful discussions with D.\ G.\ Kanhere. All the computations were performed at the cluster computing facility at HRI
(http://www.hri.res.in/cluster/)

\bibliography{phosphorene-ref} 

\end{document}


\title{\large{Peierls transition and edge reconstruction in phosphorene nanoribbons \\ Supplementary material}}

\author{Ajanta Maity}
\author{Akansha Singh}
\author{Prasenjit Sen}
\affiliation{\small{Harish-Chandra Research Institute, Chhatnag Road, Jhunsi, Allahabad 211019, India.}}  

\maketitle

\begin{figure}[h]
\scalebox{0.5}{\includegraphics{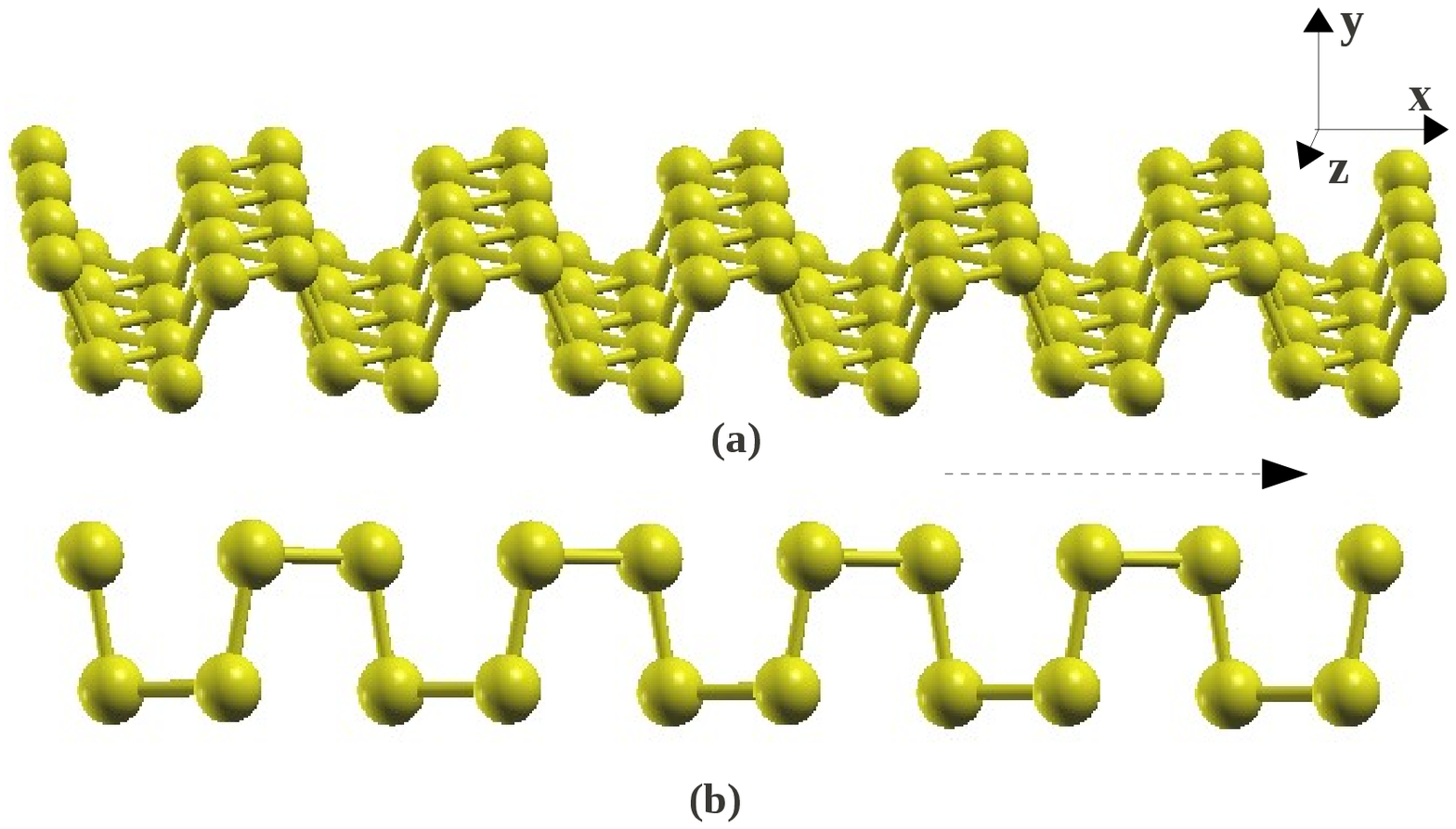}}
\caption{(Color online) (a) Structure of a r-PNR of width 4. (b) Optimized structure of a r-PNR of width 1. As is seen,
all the atoms became nearly co-planar. These PNR's are periodic along $x$ as marked by a dashed arrow.}
\end{figure}

\begin{figure}
\scalebox{0.5}{\includegraphics{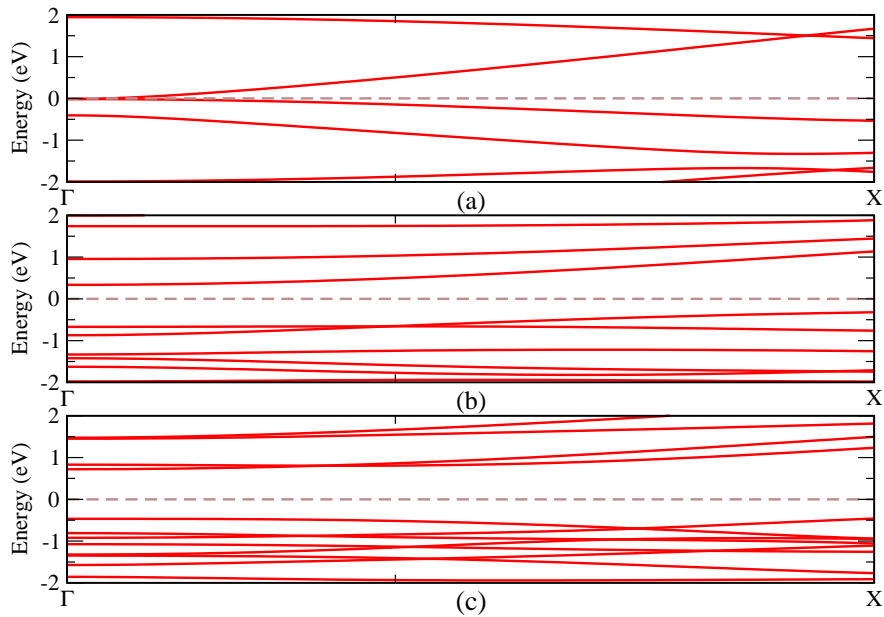}}
\caption{(Color online) Band structures of a r-PNR of (a) width 1 (b) width 2, and (c) width 3 calculated with the PBE functional.}
\end{figure}

\newpage

\begin{figure}
\scalebox{0.5}{\includegraphics{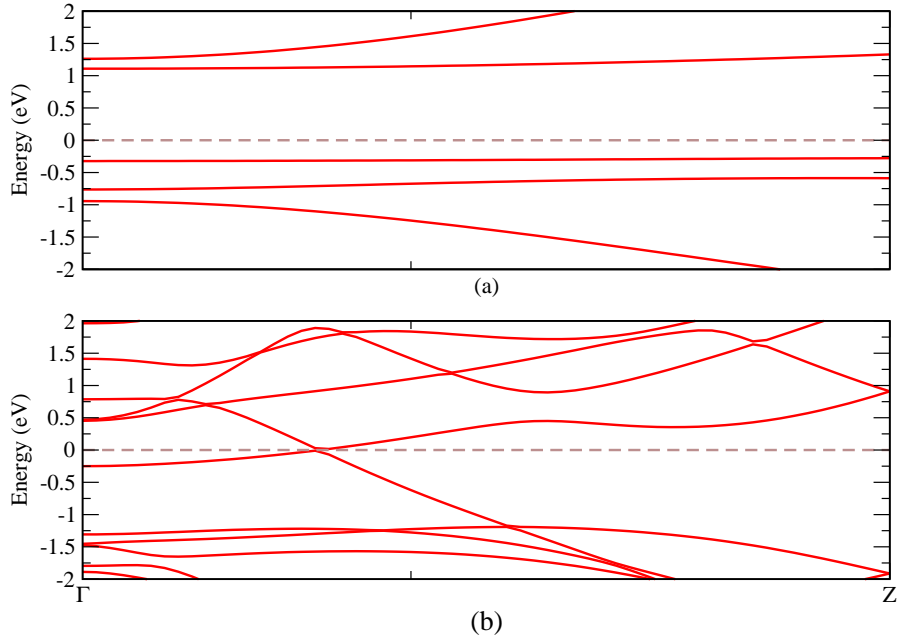}}
\caption{(Color online) Band structures of a layer-terminated l-PNR of (a) width 1 (b) width 2.}
\end{figure}

\begin{figure}
\scalebox{0.5}{\includegraphics{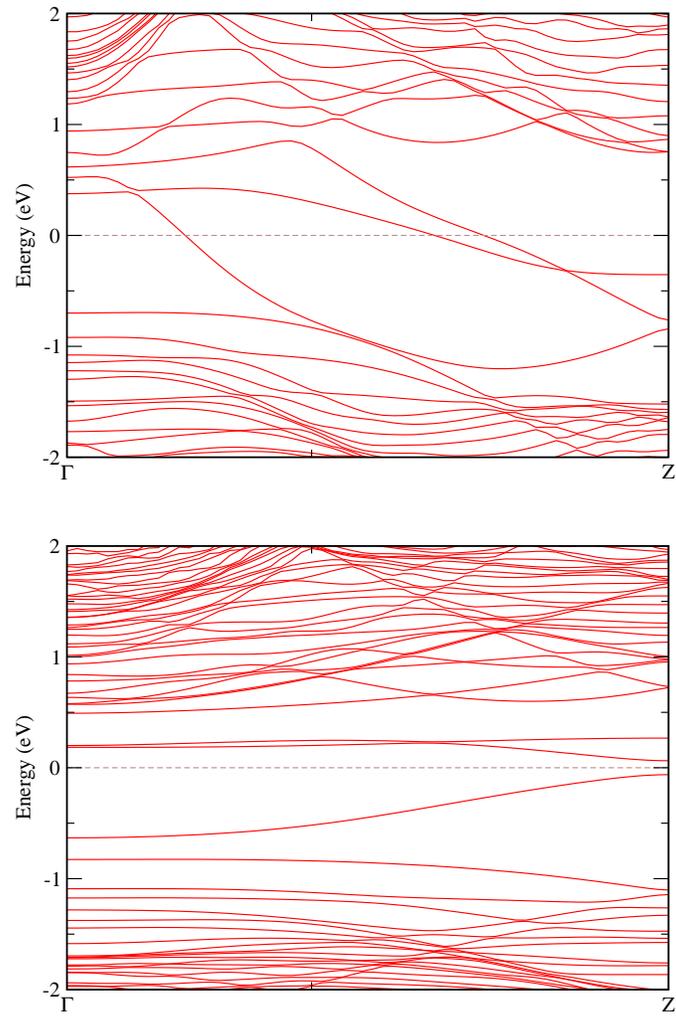}}
\caption{Band structure of a mixed PNR (a) in its layer-terminated structure, (b) after edge reconstruction and Peierls transition.}
\end{figure}


 \begin{table}[htb]
 \centering
 \caption{Band gap of pure nanoribbons of different types and widths. Gaps are calculated using PBE except for r-PNR of width 1
 as indicated.}
 \label{tab1}
 \vspace*{.25in}
  

\begin{tabular}{|c|c|c|l|}\hline

Width & r-PNR(eV)  &l-PNR(eV) & z-PNR (eV) \\\hline
    
1   & 0.45\footnote{HSE06}  &  1.38     &  metallic                 \\\hline
2    & 0.60   &  1.31   &   metallic                   \\\hline
3    & 0.99   &  1.16    & metallic                         \\\hline
4    &   0.69 &  1.11        &  0.11            \\\hline
5    &  0.51  &  1.02        &   0.15           \\\hline
6    &   0.45 &  0.97        &   0.12           \\\hline
8    &  $-$    &  0.91     &     0.11   \\ \hline
10  &   0.41 &  $-$      &       $-$   \\ \hline 
  \end{tabular}

\end{table}